\documentclass[preprint,showpacs,preprintnumbers,amsmath,amssymb]{revtex4}

\usepackage{graphicx}
\usepackage{dcolumn}
\usepackage{bm}

\begin{document}
\title{The role of the extra cellular matrix on memory}

\author{   N.Olivi-Tran(1),Sandra Kanani (2), Ian Robbins (3)}
\affiliation{(1)S.P.C.T.S., UMR-CNRS 6638,Universite de Limoges, E.N.S.C.I., 47 avenue Albert Thomas
87065 Limoges cedex, France\\
(2) IGF, 141 rue de la Cardonille, 34094 Montpellier cedex 05, France\\
(3) 	IGMM, UMR-CNRS 5535, 1919 route de Mende, 34293 Montpellier cedex 5, France}

\date{\today}

\begin{abstract}
We expose  first a biological model of memory based on one hand of the mechanical oscillations of axons during action potential and on the other hand on the changes in the extra cellular matrix composition when a mechanical strain is applied on it. Due to these changes, the stiffness of the extra cellular matrix along the most excited neurons will increase close to  these neurons due to the growth of astrocytes around them and to the elastoplastic behavior of collagen. This will create preferential paths linked to a memory effect. 
In a second part, we expose a physical model based on random walk of the action
potential on the array composed of dendrites and axons.
This last model shows that repetition of the same event leads
to long time memory of this event and that paradoxical sleep
leads to the linking of different events put into memory.
\end{abstract}
\pacs{Keywords:Memory, extra cellular matrix, random walk}
\maketitle

\section{Introduction}
Memory is one of the mystery in biology and especially in neuroscience.
Up to now no model has been able to model correctly the memory
process.
We present here two models which are in relation to explain some
features of memory. One model is based on biology of the central nervous
system and the other on physics and random walk on an array.
\section{Biological Model for Memory}
Iwasa et al \cite{iwasa} detected small lateral movement of nerve fibres during the propagation of an action potential. This lateral movement is due to swelling of the nerve fibre during excitation and has an amplitude of a few tens of  angstroms. This was an in vivo study. Moreover several in vitro studies \cite{dhill,bhill,iwasa2,tasaki} showed the same swelling and shrinking on a squid giant axon due to intracellular pressure during the action potential.

This leads to a mechanical action on the extra cellular matrix  in the nervous central system. The lateral movement due to the mechanical oscillations of the nerves or of the axons will create a strain on the surrounding extra cellular matrix. 

Let us make the hypothesis that the mechanical actions of axons (swelling and shrinking) is also present in the whole neuron during excitation but to small to be detected.  

One of the major component of the brain extra cellular matrix are glial cells. We shall focus here on a particular type of glial cells: astrocytes.  Astrocytes can feel mechanical strains which induces several different results \cite{ostrow} (Ca2+ intracellular increase,endothelin production...). Moreover, Wenzel et al. showed that high frequency stimulations of neurons caused astrocytes to wrap around the potentiated synapses \cite{wenzel}. This is the biological point of view of our model.

On the physical point of view, let us recall that another component of the extra cellular matrix is collagen type I in the central nervous system, This collagen is a biological polymer which has an elasto plastic response to stress: when a mechanical stress is applied on it and overcomes a given threshold, collagen becomes plastic and does not recover its former shape \cite{feng}. This leads to a lower concentration in collagen around the most excited neurons.

This changes in the extra cellular matrix composition along one axon or dendrite due to mechanical strain creates a preferential path. Explicitly, the difference in the extra cellular matrix composition along one neuron which is more excited compared to another will create a preferential path along the first neuron cited.

During neuronal excitations, several neurons connected one to each other via synapses will apply mechanical strains on the extra cellular matrix and therefore create a preferential path where the stiffness of the surrounding extra cellular matrix is larger due to the presence of astrocytes and the plastic behavior of collagen.

Now that preferential paths have been created in our model , let us also model the memory effect Ganguly-Fitzgerald et al \cite{ganguly} showed that for Drosophila, exposure to enriched environments (i.e. exposure to rich sensorial environments) affects the number of synapses and the size of regions involved in information processing. In our model, the preferential paths in the extra cellular matrix are regions where this last has a larger stiffness.

Therefore it will be easier for two neurons to connect along these preferential paths by a simple steric effect \cite{corey}. Indeed, if we suppose that the direction of growth of neurons (dendrites or axons) is simply leaded by the stiffness of the extra cellular matrix, the connection of two neurons via one synapse will be more frequent on preferential paths.

To conclude, for young mammals, the water concentration of brain is larger and therefore the concentration of the extra cellular matrix is lower. This will lead to vanishing preferential paths and vanishing memory.

\section{Physical Model for Memory}
Let us make the hypothesis that the preferential paths modelled in section II
exist.
We consider here the fact that memory may be represented by a random walk
of the neuronal excitation. Therefore, the statistical approach
that we give here and which is in relation with the preferential paths
will give the probability for each given event to be put into memory.
In this section, we consider only the preferential paths which
we consider already built. We do not consider here the effect
of the extra cellular matrix.

We model the assembly of neurons by a packing of overlapping dendrites and axons
 at first approximation. This random array without excluding volume may be
modelled by a simple random walk or Brownian motion.

Let us consider the Brownian motion of a particle (equivalent to the path of a neuronal excitation)  and let us try to find it
at location $\bf R$ at time $t$, knowing that it is at the origin
at time $t=0$.
The central limit theorem allows one to obtain the distribution $R_N$
when $N$ tends towards infinity. This central limit theorem may be written
:
\begin{equation}
P({\bf R},t) \rightarrow \frac{1}{(4 \pi D t)^{d/4}}\exp -\frac{\bf R^2}{4Dt}
\end{equation}
with $t = N \tau$ and the diffusion constant $D=<{\bf r^2}>/2 \tau$
Let us now examine the scalar Brownian motion.
If $g$ is the gain at time $t$ knowing that it was zero at time $t=0$,
the probability to have a gain $g>0$ or a loss $g<0$ after a time $t$
(the time between two events is then equivalent to the unit interval) is:
\begin{equation}
P(g,t)=\frac{1}{(2 \pi t)^{3/4}} \exp{-\frac{g^2}{2t}}
\end{equation}
After one neuronal excitation, this last equation may be related to the length
of the neuronal path by replacing the gain $g$ by the end to end length of this path.
Note that their may be crossings of the path which model the synapses.
The probability to have exactly the same path with the same gain $g$
but in the inverse direction is then exactly the same as equation (2).
If the path is long i.e. $t$ is large the probability will tend to zero.
This is a model for immediate memory.
But for long time memory, when exactly the same neuronal excitation is repeated
$n$ times independently, the probability becomes:
\begin{equation}
nP(g,t)=n\frac{1}{(2 \pi t)^{3/4}} \exp{-\frac{g^2}{2t}}
\end{equation}
with a maximum equal to one;
and the inverse path (long time memory) has the same probability.

All this first part of our model is valid if the paths have an underlying random array for the random walk. This underlying array is simply the neurons
with their dendrites and axons. Once a random walker (or neuronal excitation)
jumps from one axon (or dendrite) to another axon (or dendrite) it creates
a synapse which remains due to the action of astrocytes (see section II).
Moreover, the extracellular matrix which is compressed in the neighbouring
of the excited neuron will create a void path in the collagen leading
to less mechanical resistance and leading to preferential  growth of another dendrite or axon on
the same path.

Let us known analyze the action of paradoxical sleep.
We model the paradoxical sleep as a random walker which will visit
all the underlying lattice, i.e. all the existing paths will be visited
by the neuronal excitation.
For that we need to take a gain $g$ equal to zero in order
to model the return to the origin.
The return to the origin (null gain) follows the law:
\begin{equation}
P(0,t)=\frac{1}{(2 \pi t)^{3/4}}
\end{equation}
Following the duration of paradoxical sleep, one
 has to calculate:
\begin{equation}
N_0=\int_O^t P(0,t')dt'=\int_0^t \frac{dt'}{(2 \pi t')^{3/4}}=\frac{2}{3}\pi t^{1/4}
\end{equation}
Therefore, the probability to link to different long memory paths is equal
to $\frac{2}{3}\pi t^{1/4}$.
This may explain that after paradoxical sleep, two different events which
are memorized may be linked.

\section{Conclusion}
We made two models of memory, one biological based on bibliographic data,
the other based on random walks on the dendrites and axons formed array.
The biological model shows that the action of the extracellular matrix
is essential for memory. The physical model shows that the probability
to memorize an event is dependent on the number of times this event has excited
the corresponding neurons.

\end{document}